\documentclass[aps,prl,preprint]{revtex4}

\usepackage{epsfig}

\draft

\begin{document}
\author{Hong-shi Zong$^{1,2,3}$, Deng-ke He$^{1}$, Feng-yao Hou$^{1}$, and Wei-min Sun$^{1,2}$}
\address {$^{1}$ Department of Physics, Nanjing University, Nanjing 210093, China}
\address {$^{2}$ Joint Center for Particle, Nuclear Physics and Cosmology, Nanjing 210093, China}
\address{$^{3}$ CCAST(World Laboratory), P.O. Box 8730, Beijing 100080, China}

\title{General formula for the four-quark condensate and 
      vacuum factorization assumption}

\begin{abstract}
By differentiating the dressed quark propagator with respect to a variable background field, the linear response of the dressed quark propagator in the presence of the background field can be obtained. From this general method, using the vector background field as an illustration, we derive a general formula for the four-quark condensate $\langle\tilde{0}|:\bar{q}(0)\gamma_{\mu}q(0)\bar{q}(0)\gamma_{\mu}q(0):|\tilde{0}\rangle$. This formula contains the corresponding fully dressed vector vertex and it is shown that  factorization for $\langle\tilde{0}|:\bar{q}(0)\gamma_{\mu}q(0)\bar{q}(0)\gamma_{\mu}q(0):|\tilde{0}\rangle$ holds only when the dressed vertex is taken to be the bare one. This property also holds for all other type of
four-quark condensate.
 
\bigskip


\bigskip

E-mail: zonghs@chenwang.nju.edu.cn, nuclphys@nju.edu.cn (H.-s. Zong)

\bigskip


\end{abstract}

\maketitle

It is well-known that one of the central problems of strong-interaction physics is in understanding the structure of the ground state of quantum Quantum Chromodynamics, the QCD vacuum. One way to characterize this structure is by means of various vacuum condensates, such as the two-quark condensate, the gluon condensate and the four-quark condensate, etc. These condensates are essential for describing the strong interaction physics using the QCD sum rule method [1,2]. Clearly, an important ingredient of the QCD sum rule method is to determine the various vacuum condensates as precisely as possible. In the numerical 
predictions of the QCD sum rule method, the largest uncertainties reside in the actual values for the four-quark condensates. They remained a matter of constant debate over the last decades [3-11]. One central point in these discussions is the question whether a four-quark condensate can be factorized more or less accurately into a product of two-quark condensates. The reason for this is that at present there does not exist a general approach for calculating this quantity. In this paper we try to present such a general approach and use it to analyse the factorization problem of the four-quark condensate.   
 
In order to make this paper self-contained, let us first recall the definition of vacuum condensates in the QCD sum rule formalism. In the QCD sum rule, one often postulates that quark propagators are modified by the long-range confinement part of QCD; but the modification is soft in the sense that at short distances the difference between the exact and perturbative propagators vanishes. To formalize this statement, one can write the ``exact'' propagator $G(x)$ as the vacuum expectation value of the following T-product in the ``exact'' QCD vacuum $|\tilde{0}\rangle$:
\begin{equation}
G_{ij}(x,y)\equiv\langle\tilde{0}|T[q_i(x)\bar{q}_j(y)]|\tilde{0}\rangle.
\end{equation}
According to the spirit of Wick theorem, one formally writes the T-product as the sum
\begin{equation}
T[q_i(x)\bar{q}_j(y)]\equiv\underbrace{q_i(x)\bar{q}_j(y)}+:q_i(x)\bar{q}_j(y):
\end{equation}
of the ``pairing'' and the ``normal'' product. The ``pairing'' is just the expectation value of the T-product over the perturbative QCD vacuum $|0\rangle$
\begin{equation}
\underbrace{q_i(x)\bar{q}_j(y)}\equiv\langle 0|T[q_i(x)\bar{q}_j(y)]|0\rangle\equiv G_{ij}^{pert}(x,y),
\end{equation}
i.e., the perturbative propagator. Here we emphasize that  Wick theorem is obtained in free quantum field theories and there is little reason to expect that it is still of validity in the case of exact QCD vacuum (since the vacuum $|\tilde{0}\rangle$ is highly nontrivial, it is not clear how to define the normal product of field operators in this case), therefore Eq. (2) should be understood as a formal definition of the ``normal'' product. By this definition, we have the following two-quark vacuum condensate $\langle\tilde{0}|:\bar{q}(x)q(y):|\tilde{0}\rangle$:
\begin{eqnarray}
&&\langle \tilde{0}|:\bar{q}_i(x)q_j(y):|\tilde{0}\rangle\equiv
\langle\tilde{0}|T[\bar{q}_i(x)q_j(y)]|\tilde{0}\rangle
-\langle 0|T[\bar{q}_i(x)q_j(y)]|0\rangle\nonumber \\
&&=(-)\left[G_{ji}(y,x)-G_{ji}^{pert}(y,x)\right]
=(-)\int\frac{d^4 q}{(2\pi)^4}e^{i q\cdot (y-x)}\left[G(q^2)-G^{pert}(q^2)\right]_{ji}.
\end{eqnarray}
Here we would like to stress again that the bracketting colons $:~:$ is only a notation which means that we subtract the contribution of the perturbative term $G^{pert}(x)$ from $G(x)$ [12,13]. Thus, our assumption $\langle\tilde{0}|:\bar{q}q:|\tilde{0}\rangle$ $\not=0$ is equivalent to the statement $G(x)\not= G^{pert}(x)$. 

Similarly, we have the four-quark vacuum condensate:
\begin{eqnarray}
&& \langle \tilde{0}|:\bar{q}(x)\Lambda^{(1)}q(x)\bar{q}(y)\Lambda^{(2)}q(y):|\tilde{0}\rangle \nonumber\\
&&\equiv\langle \tilde{0}|T\left[\bar{q}(x)\Lambda^{(1)}q(x)
\bar{q}(y)\Lambda^{(2)}q(y)\right]|\tilde{0}\rangle
-\langle 0|T\left[\bar{q}(x)\Lambda^{(1)}q(x)
\bar{q}(y)\Lambda^{(2)}q(y)\right]|0\rangle \nonumber\\
&&-\langle \tilde{0}|:\bar{q}(x)\Lambda^{(1)}\underbrace{q(x)
\bar{q}(y)}\Lambda^{(2)}q(y):|\tilde{0}\rangle
-\langle \tilde{0}|:\underbrace{\bar{q}(x)\Lambda^{(1)}q(x)
\bar{q}(y)\Lambda^{(2)}q(y)}:|\tilde{0}\rangle\\
&&-\langle \tilde{0}|:\underbrace{\bar{q}(x)\Lambda^{(1)}q(x)}
\bar{q}(y)\Lambda^{(2)}q(y):|\tilde{0}\rangle
-\langle \tilde{0}|:\bar{q}(x)\Lambda^{(1)}q(x)
\underbrace{\bar{q}(y)\Lambda^{(2)}q(y)}:|\tilde{0}\rangle\nonumber,
\end{eqnarray}
here the $\Lambda^{(i)}$ stands for a matrix in Dirac and color space. 

To be more concrete, in the following let us take the case $\Lambda^{(1)}=\Lambda^{(2)}=\gamma_{\mu}
\otimes I_C$ (where $I_C$ denotes the unit matrix in the color space) as an example to illustrate the general approach for calculating the four-quark condensate. When $\Lambda^{(1)}=\Lambda^{(2)}=\gamma_{\mu}\otimes I_C$, it can be shown that the last four terms on the right hand side of Eq. (5) vanish. For instance, for the third term, one can write
\begin{eqnarray}
&&\langle \tilde{0}|:\bar{q}(x)\Lambda^{(1)}\underbrace{q(x)
\bar{q}(y)}\Lambda^{(2)}q(y):|\tilde{0}\rangle
=\langle \tilde{0}|:\bar{q}_i(x)\Lambda^{(1)}_{ij}\underbrace{q_{j}(x)\bar{q}_{m}(y)}\Lambda^{(2)}_{mn}q_{n}(y):|\tilde{0}\rangle \nonumber\\
&&=\Lambda^{(1)}_{ij} G^{pert}_{jm}(x,y)\Lambda^{(2)}_{mn} 
 \langle \tilde{0} |:\bar{q}_i(x) q_n (y) :|\tilde{0}\rangle
=\frac{1}{12}\Lambda^{(1)}_{ij} G^{pert}_{jm}(x,y)\Lambda^{(2)}_{mn}\delta_{in}\langle \tilde{0} |:\bar{q}(x) q (y) :|\tilde{0}\rangle \nonumber\\
&&=\frac{1}{12}Tr_{\gamma C}[\Lambda^{(1)}G^{pert}(x,y)\Lambda^{(2)}] \langle \tilde{0} |:\bar{q}(x) q (y) :|\tilde{0}\rangle, 
\end{eqnarray}
where the trace operation is on the color and Dirac indices. 
For $\Lambda^{(1)}=\Lambda^{(2)}=\gamma_{\mu}\otimes I_C$, one has
\[ Tr_{\gamma C}[\gamma_{\mu}\otimes I_C G^{pert} \gamma_{\mu}\otimes I_C ]
=4 Tr_{\gamma C}[G^{pert}]=0,\]
where we have assumed that we are working in the chiral limit. It can be shown that, in the chiral limit, the scalar self-energy function of the perturbative quark propagator vanishes to all orders of perturbation theory [14], hence this term vanishes. For the fifth term, one can write
\begin{eqnarray}
&&\langle \tilde{0}|:\underbrace{\bar{q}(x)\Lambda^{(1)}q(x)}
\bar{q}(y)\Lambda^{(2)}q(y):|\tilde{0}\rangle\nonumber\\
&&=\langle \tilde{0}|:\underbrace{\bar{q}_{i}(x)\Lambda^{(1)}_{ij}q_{j}(x)}
\bar{q}_{m}(y)\Lambda^{(2)}_{mn}q_{n}(y):|\tilde{0}\rangle=-G^{pert}_{ji}(x,x)\Lambda^{(1)}_{ij}\Lambda^{(2)}_{mn} 
\langle \tilde{0} |:\bar{q}_{m}(y) q_{n}(y) :|\tilde{0} \rangle\nonumber\\
&&=-\frac{1}{12}Tr_{\gamma C}[G^{pert}(x,x)\Lambda^{(1)}]
Tr_{\gamma C}[\Lambda^{(2)}]\langle \tilde{0} |:\bar{q}(y) 
q_(y) : | \tilde{0} \rangle.
\end{eqnarray}
For $\Lambda^{(1)}=\Lambda^{(2)}=\gamma_{\mu}\otimes I_C$, one has $Tr_{\gamma C}[\Lambda^{(2)}]=0$, therefore this term vanishes. By similar procedures, one can prove that the other two terms also vanish. Therefore one has
\begin{eqnarray}
&&\langle\tilde{0}|:\bar{q}(x)\gamma_{\mu}q(x)\bar{q}(y)\gamma_{\mu}q(y):|\tilde{0}\rangle \nonumber\\
&&=\langle\tilde{0}|T\left[\bar{q}(x)\gamma_{\mu}q(x)
\bar{q}(y)\gamma_{\mu}q(y)\right]|\tilde{0}\rangle
-\langle 0|T\left[\bar{q}(x)\gamma_{\mu}q(x)
\bar{q}(y)\gamma_{\mu}q(y)\right]|0\rangle.
\end{eqnarray}
As can be seen from Eq. (8), in order to calculate the four-quark condensate, one must know how to calculate $\langle\tilde{0}|T\left[\bar{q}(x)\gamma_{\mu}q(x)
\bar{q}(y)\gamma_{\mu}q(y)\right]|\tilde{0}\rangle$ and $\langle 0|T\left[\bar{q}(x)\gamma_{\mu}q(x)
\bar{q}(y)\gamma_{\mu}q(y)\right]|0\rangle$ in advance. The main task now is to determine the vacuum expectation value (VEV) of the above T-product of four quark field operators in a consistent way. In the following we shall derive a general formula for these VEV using the background field method. 

Let us now study the linear response of the dressed quark propagator in the presence of the variable background field.  The presence of the variable vector background  field implies that the dressed quark propagator ${\cal{G}}[{\cal{V}}](x)$ (in the presence of the background vector field ${\cal{V}}_{\mu}(x)$) is evaluated with an additional term $\Delta S\equiv\int d^4x~\bar{q}(x)i\gamma_{\mu}q(x){\cal{V}}_{\mu}(x)$ added to the usual QCD action. In the Euclidean space and the chiral limit, it can be written as
\begin{eqnarray}
{\cal{G}}_{ij}[{\cal{V}}](x)&=&
\frac{\int{\cal{D}}\bar{q} 
{\cal{D}}q{\cal{D}}A~ q_i(x)\bar{q}_j(0)\exp\left\{-S[\bar
{q},q,A]+\int d^4x~\bar{q}(x)i\gamma_{\mu}q(x){\cal{V}}_
{\mu}(x)\right\}}{\int{\cal{D}}\bar{q}{\cal{D}}q{\cal{D}}A 
\exp\left\{-S[\bar{q},q,A]+\int d^4x~\bar{q}(x)i\gamma_{\mu}
q(x){\cal{V}}_{\mu}(x)\right\}},
\end{eqnarray}
where
\begin{equation}
S[\bar{q},q,A]=\int d^4x\left\{\bar{q}({\not\!\!{\partial}}-ig\frac{\lambda^{a}}{2}{\not\!\!{A^a}})q+\frac{1}{4}F^{a}_{\mu\nu}F^{a}_{\mu\nu}\right\},
\end{equation}
and $F^{a}_{\mu\nu}=\partial_{\mu}A^{a}_{\nu}-\partial_{\nu}A^{a}_{\mu}+gf^{abc}A^{b}_{\mu}A^{c}_{\nu}$. We leave the gauge fixing term, the ghost field term and its integration measure to be understood. In this paper we adopt the following conventions: The Dirac matrices are Hermitian and satisfy the algebra $[\gamma_{\mu},\gamma_{\nu}]=2\delta_{\mu\nu}$ and $\gamma_5$ is defined as $\gamma_5\equiv \gamma_1\gamma_2\gamma_3\gamma_4$ so that $Tr_{\gamma}[\gamma_5\gamma_{\mu}\gamma_{\nu}\gamma_{\rho}\gamma_{\sigma}]=4\epsilon_{\mu\nu\rho\sigma}$, $\epsilon_{1234}=1$.

If one focuses only on the linear response term of ${\cal{G}}[{\cal{V}}](x)$, we have
\begin{eqnarray}
{\cal{G}}_{ij}[{\cal{V}}](x)
&=&\frac{\int{\cal{D}}\bar{q}{\cal{D}}q{\cal{D}}A~q_i(x)\bar{q}_j(0)\left[1+\int d^4y~\bar{q}(y)i\gamma_{\mu}q(y){\cal{V}}_{\mu}(y)+\cdot\cdot\cdot\right]\exp\{-S[\bar{q},q,A]\}}
{\int{\cal{D}}\bar{q}{\cal{D}}q{\cal{D}}A\left[1+\int d^4y\bar{q}(y)i\gamma_{\mu}q(y){\cal{V}}_{\mu}(y)+\cdot\cdot\cdot\right]\exp\{-S[\bar{q},q,A]\}}\nonumber \\
&=&\frac{\langle\tilde{0}|T[q_i(x)\bar{q}_j(0)]|\tilde{0}   \rangle+ \int d^4y\langle\tilde{0}|T[q_i(x)\bar{q}_j(0)\bar
{q}(y)i\gamma_{\mu}q(y){\cal{V}}_{\mu}(y)]|\tilde{0}\rangle+\cdot\cdot\cdot}
{1+\int d^4y\langle\tilde{0}|T[\bar{q}(y)i\gamma_{\mu}q(y)
{\cal{V}}_{\mu}(y)]|\tilde{0}\rangle+\cdot\cdot\cdot} 
\nonumber \\
&=&\langle\tilde{0}|T[q_i(x)\bar{q}_j(0)]|\tilde{0}\rangle
 +\int d^4y\langle\tilde{0}|T[q_i(x)\bar{q}_j(0)\bar{q}(y)i\gamma_{\mu}q(y){\cal{V}}_{\mu}(y)]|\tilde{0}\rangle \nonumber \\
&& -\langle\tilde{0}|T[q_i(x)\bar{q}_j(0)]|\tilde{0}\rangle
\int d^4y\langle\tilde{0}|T[\bar{q}(y)i\gamma_{\mu}q(y)
{\cal{V}}_{\mu}(y)]|\tilde{0}\rangle+\cdot\cdot\cdot.
\end{eqnarray}
The third term on the right hand side of the last line of Eq. (11) vanishes since $\langle\tilde{0}|T[\bar{q}(y)i\gamma_{\mu}q(y)]|\tilde{0}\rangle=0$ (since the vacuum state is Lorentz invariant, the vacuum expectation value
of any local operator which carries a four-vector index vanishes identically), so
\begin{eqnarray}
{\cal{G}}_{ij}^{{\cal{V}}}(x)\equiv
\int d^4y\langle\tilde{0}|T[q_i(x)\bar{q}_j(0)\bar{q}(y)i\gamma_{\mu}q(y){\cal{V}}_{\mu}(y)]|\tilde{0}\rangle 
\end{eqnarray}
is the linear response term of the exact quark propagator in the presence of the vector background field. 
At this point, one remark is needed. The above background 
field method is general and one can study the linear 
response of the dressed quark propagator in the presence of
background field of any other type. In all cases, up to first order of the background field, the expression for the dressed quark propagator contains three terms, similar to the right hand side of the last line of Eq. (11). In the case of the tensor, axial vector or pseudoscalar background field, using similar arguments as the vector case, one can show that the linear response term of the dressed quark propagator is given by an expression similar to (12). However, in the case of the scalar background field, the $\langle {\tilde 0}|T[\bar{q}(y)q(y)]|{\tilde 0}\rangle$ does not vanish and the linear response term of the dressed quark propagator consists of two terms:
\begin{eqnarray}
{\cal{G}}_{ij}^{{\cal{S}}}(x)&=&
-\int d^4y\langle\tilde{0}|T[q_i(x)\bar{q}_j(0)\bar{q}(y)q(y){\cal{S}}(y)]|\tilde{0}\rangle\nonumber\\
&&+\langle\tilde{0}|T[q_i(x)\bar{q}_j(0)]|\tilde{0}\rangle\int d^4y\langle\tilde{0}|T[\bar{q}(y)q(y){\cal{S}}(y)]|\tilde{0}\rangle,
\end{eqnarray}  
where ${\cal{S}}(y)$ denotes the variable scalar background  field. As will be shown below, the contribution of the second term on the right hand side of Eq. (13) should be 
taken into account in the calculation of the four-quark condensate $\langle {\tilde 0}|:{\bar q}(0)q(0){\bar q}(0)
q(0):|{\tilde 0} \rangle$. 

Now let us turn back to the discussion of the vector background field case. In order to obtain a exact expression for $\int d^4y\langle\tilde{0}|T[q(x)\bar{q}(0)\bar{q}(y)i\gamma_{\mu}q(y){\cal{V}}_{\mu}(y)]|\tilde{0}\rangle$, we expand the dressed quark propagator ${\cal{G}}^{-1}[{\cal{V}}]$ in powers of ${\cal{V}}$ as follows
\begin{equation}
{\cal{G}}^{-1}[{\cal{V}}]={\cal{G}}^{-1}[{\cal{V}}]\mid_{{\cal{V}}_{\mu}=0}+\frac{\delta {\cal{G}}^{-1}[{\cal{V}}]}{\delta {\cal{V}}}
\mid_{{\cal{V}}_{\mu}=0}{\cal{V}}+\cdot\cdot\cdot\equiv G^{-1}+{\cal{V}}_{\mu}\Gamma_{\mu}+\cdot\cdot\cdot,
\end{equation}
which leads to the following formal expansion
\begin{equation}
{\cal{G}}[{\cal{V}}]=G-G{\cal{V}}_{\mu}\Gamma_{\mu}G+\cdot\cdot\cdot,
\end{equation}
with
\begin{equation}
\Gamma_{\mu}(y_1,y_2;z)=\left[\frac{\delta {\cal{G}}^{-1}[{\cal{V}}](y_1,y_2)}{\delta {\cal{V}}_{\mu}(z)}\right]_{{\cal{V}}_{\mu}=0}.
\end{equation}
In coordinate space the dressed vector vertex $\Gamma_{\mu}(x,y;z)$ is given as the functional derivative of the inverse quark propagator ${\cal{G}}^{-1}[{\cal{V}}]$ with respect to the variable background field ${\cal{V}}_{\mu}(x)$. Note that Eq. (15) holds for both the exact quark propagator and the perturbative one.

Eq. (15) is a compact notation and its explicit form reads 
\begin{eqnarray}
{\cal{G}}[{\cal{V}}](y_1,y_2)&=&G(y_1,y_2)-\int d^4 u_1 d^4 u_2 d^4 z~G(y_1,u_1)\Gamma_{\mu}(u_1,u_2;z)\cdot {\cal{V}}_{\mu}(z)G(u_2,y_2)
+\cdot\cdot\cdot\nonumber\\
&=&G(y_1,y_2)-\int d^4z\int \frac{d^4q}{(2\pi)^4}\int \frac{d^4P}{(2\pi)^4}~e^{-i(P+\frac{q}{2})\cdot y_1}e^{i(P-\frac{q}{2})\cdot y_2}e^{i q\cdot z}\nonumber\\
&&\times G(P+\frac{q}{2})\Gamma_{\mu}(P;q)\cdot {\cal{V}}_{\mu}(z)G(p-\frac{q}{2})+\cdot\cdot\cdot,
\end{eqnarray}
where $P$ and $q$ are the relative and total momentum of the quark-antiquark pair, respectively.

Comparing Eq. (11) with (17), we obtain the linear response term of the exact quark propagator ${\cal{G}}^{{\cal{V}}}(x)$ in the presence of variable background field
\begin{eqnarray}
&&{\cal{G}}_{ij}^{{\cal{V}}}(x)\equiv\int d^4y\langle\tilde{0}|T[q_i(x)\bar{q}_j(0)\bar{q}(y)i\gamma_{\mu}q(y){\cal{V}}_{\mu}(y)]|\tilde{0}\rangle\\
&&=-\int d^4z\int \frac{d^4q}{(2\pi)^4}\int \frac{d^4P}{(2\pi)^4}~e^{-i(P+\frac{q}{2})\cdot x}e^{i q\cdot z}\left[G(P+\frac{q}{2})\Gamma_{\mu}(P;q)\cdot {\cal{V}}_{\mu}(z)G(P-\frac{q}{2})\right]_{ij}\nonumber.
\end{eqnarray}
Setting $x=0$ and taking ${\cal{V}}_{\mu}(y)=V_{\mu}\delta(y)$ ($V_{\mu}$ is an constant vector external field) in 
Eq. (18), and then multiplying by $(\gamma_{\mu})_{ji}$,
we have
\begin{eqnarray}
&&\langle\tilde{0}|T\left[\bar{q}(0)\gamma_{\mu}q(0)\bar{q}(0)\gamma_{\mu}q(0)\right]|\tilde{0}\rangle\nonumber\\
&&=(-i)\int\frac{d^4q}{(2\pi)^4}\int \frac{d^4P}{(2\pi)^4}~Tr_{\gamma C}\left[\gamma_{\mu}G(P+\frac{q}{2})\Gamma_{\mu}(P;q)
G(P-\frac{q}{2})\right].
\end{eqnarray}

Similarly, we have 
\begin{eqnarray}
&&\langle 0|T\left[\bar{q}(0)\gamma_{\mu}q(0)\bar{q}(0)\gamma_{\mu}q(0)\right]|0\rangle\nonumber\\
&&=(-i)\int\frac{d^4q}{(2\pi)^4}\int \frac{d^4P}{(2\pi)^4}~Tr_{\gamma C}\left[\gamma_{\mu}G^{pert}(P+\frac{q}{2})\Gamma^{pert}_{\mu}(P;q)G^{pert}(P-\frac{q}{2})\right].
\end{eqnarray}
where $\Gamma^{pert}_{\mu}(P;q)$ is the corresponding perturbative vector vertex.

Putting Eqs. (19) and (20) into Eq. (8), one obtains the main result of this paper, namely the general formula for the four-quark condensate
\begin{eqnarray}
&&\langle\tilde{0}|:\bar{q}(0)\gamma_{\mu}q(0)\bar{q}(0)\gamma_{\mu}q(0):|\tilde{0}\rangle\nonumber\\
&&=(-i)\left\{\int\frac{d^4q}{(2\pi)^4}\int \frac{d^4P}{(2\pi)^4}~Tr_{\gamma C}\left[\gamma_{\mu}G(P+\frac{q}{2})\Gamma_{\mu}(P;q)G(P-\frac{q}{2})\right]\right.\nonumber\\
&&\left.-\int\frac{d^4q}{(2\pi)^4}\int \frac{d^4P}{(2\pi)^4}~Tr_{\gamma C}\left[\gamma_{\mu}G^{pert}(P+\frac{q}{2})\Gamma^{pert}_{\mu}(P;q)G^{pert}(P-\frac{q}{2})\right]\right\}.
\end{eqnarray}
At this point, it is interesting to compare Eq. (21) with the general formula for the vector vacuum susceptibility $\chi^{{\cal{V}}}$ [15]:
\begin{eqnarray}
&&\chi^{{\cal{V}}}\propto\int \frac{d^4P}{(2\pi)^4}~Tr_{\gamma C}\left\{\gamma_{\mu}G(P)\Gamma_{\mu}(P;0)G(P)-\gamma_{\mu}G^{pert}(P)\Gamma^{pert}_{\mu}(P;0)G^{pert}(P)\right\}.
\end{eqnarray}
In the QCD sum rule external field method, one often introduces various vacuum susceptibilities to characterize the properties of the QCD vacuum. It is generally believed that the vacuum susceptibilities and the vacuum condensates are independent quantities. However, comparing 
Eqs. (21) and (22), it is easy to see that there exists some intrinsic relations between the two quantities 
$\langle\tilde{0}|:\bar{q}(0)\gamma_{\mu}q(0)\bar{q}(0)\gamma_{\mu}q(0):|\tilde{0}\rangle$ and $\chi^{{\cal{V}}}$. 

The whole procedures above are presented in the unrenormalized languages. Once the technique for deriving
the four-quark condensate is known one can easily repeat the whole derivation in the renormalized languages.
Essentially one only needs to modify the action of the theory to include the necessary counterterms and repeat the above procedure (more details can be found in Refs. [16,17]). 

Now let us turn back to formula (21). We note that formula (21) is formally model-independent,
and due to presence of the subtraction term (the second term on the right hand side of Eq. (21)) the four-quark condensate is free of ultraviolet divergences, since in the large momentum region $G(p)$ and $\Gamma_{\mu}(k;P)$ coincide with $G^{pert}(p)$ and $\Gamma_{\mu}^{pert}(k;P)$, respectively. However, in order to apply this formula, one should know the values of the quark propagators $G(p)$, $G^{pert}(p)$ and the vertex functions $\Gamma_{\mu}(k;P)$, $\Gamma_{\mu}^{pert}(k;P)$ for the whole momentum region in advance. At present it is very difficult to calculate the above functions for the whole momentum region from first principles of QCD. Thus in practical calculation of the four-quark condensates one has to resort to nonperturbative QCD model. Over the past few years, considerable progress has been made in the framework of the rainbow-ladder approximation of the Dyson-Schwinger (DS) approach [16-18], which provides a successful description of various nonperturbative aspects of strong interaction physics. Just as was pointed out in Refs. [15,19,20], the DS approach provides a general framework for calculating the four functions $G(p)$, $G^{pert}(p)$, $\Gamma_{\mu}(k;P)$ and $\Gamma_{\mu}^{pert}(k;P)$. For details we refer the readers to the above references. 

In the above we have derived the general formula (21) for
the four-quark condensate $\langle\tilde{0}|:\bar{q}(0)\gamma_{\mu}q(0)\bar{q}(0)\gamma_{\mu}q(0):|\tilde{0}
\rangle$. One can use the same procedure to derive the general formula for the four-quark condensate of any other type. For example, for the four-quark condensate $\langle\tilde{0}|:\bar{q}(0)q(0)\bar{q}(0)q(0):|\tilde{0}\rangle$,
we have the following formula
\begin{eqnarray}
&&\langle\tilde{0}|:\bar{q}(0)q(0)\bar{q}(0)q(0):|\tilde{0}\rangle-[\langle\tilde{0}|:\bar{q}(0)q(0):|\tilde{0}
\rangle]^2\nonumber\\
&&=(-i)\left\{\int\frac{d^4q}{(2\pi)^4}\int \frac{d^4P}{(2\pi)^4}~Tr_{\gamma C}\left[G(P+\frac{q}{2})\Gamma(P;q)G(P-\frac{q}{2})\right]\right.\nonumber\\
&&\left.-\int\frac{d^4q}{(2\pi)^4}\int \frac{d^4P}{(2\pi)^4}~Tr_{\gamma C}\left[G^{pert}(P+\frac{q}{2})\Gamma^{pert}(P;q)G^{pert}(P-\frac{q}{2})\right]\right\},
\end{eqnarray}
where $\Gamma(P;q)$ and $\Gamma^{pert}(P;q)$ denote the corresponding exact and perturbative dressed scalar vertex, 
respectively. As was noted before, the linear response term  ${\cal{G}}_{ij}^{{\cal{S}}}(x)$ consists of two terms (see
Eq. (13)), and the contribution of the second term in Eq. (13) leads to the appearance of the extra term $-[\langle\tilde{0}|:\bar{q}(0)q(0):|\tilde{0}\rangle]^2$ 
on the left hand side of Eq. (23).

As was said above, one central point in the discussion on the four-quark condensate is the question whether a four-quark condensate can be factorized more or less accurately into a product of two-quark condensates. Now we have obtained a closed formula for a general type of four-quark condensate, we can make use of it to analyse the factorization problem. If one adopts the vacuum saturation assumption, one has  
\begin{equation}
\langle \tilde{0}|:\bar{q}\Lambda^{(1)}q\bar{q}\Lambda^{(2)}q:|\tilde{0}\rangle\approx\frac{1}{12^2}\left[Tr_{\gamma C}\Lambda^{(1)}Tr_{\gamma C}\Lambda^{(2)}-Tr_{\gamma C}\left(\Lambda^{(1)}\Lambda^{(2)}\right)\right]\left[\langle \tilde{0}|:\bar{q}q:|\tilde{0}\rangle\right]^2.
\end{equation}
i.e., the four-quark condensate is factorized into a product of two-quark condensates. In the following we shall still use the four-quark condensate $\langle\tilde{0}|:\bar{q}(0)\gamma_{\mu}q(0)\bar{q}(0)\gamma_{\mu}q(0):|\tilde{0}\rangle$ as an illustration to analyze this problem. We note that the two-quark condensate $\langle \tilde{0}|:\bar{q}q:|\tilde{0}\rangle$ is related to the two-point correlation function, i.e., the quark propagator, while it can be seen from Eq. (21) that 
besides the two-point correlation function, the three-point correlation function, i.e., the fully dressed vertex 
$\Gamma_{\mu}$, also enters into the expression for the four-quark condensate $\langle\tilde{0}|:\bar{q}(0)\gamma_{\mu}q(0)\bar{q}(0)\gamma_{\mu}q(0):|\tilde{0}\rangle$. Therefore, factorization of the four-quark condensate is essentially equivalent to neglecting the influences of the corresponding dressed vertex (three-point correlation function). The simplest approximation scheme for the dressed vertex is the bare vertex approximation: 
$\Gamma_{\mu}(P;q)\approx -i \gamma_{\mu}$, $\Gamma_{\mu}^{pert}(P;q)\approx -i \gamma_{\mu}$. Interestingly, it can be shown that only under this approximation does the factorization of the four-quark condensate hold. The following is the proof of this assertion.

In the case of bare vertex approximation, the first term on the right hand side of Eq. (21) can be written as
\begin{eqnarray}
&&(-)\int\frac{d^4q}{(2\pi)^4}\int \frac{d^4P}{(2\pi)^4}~Tr_{\gamma C}\left[\gamma_{\mu}G(P+\frac{q}{2})\gamma_{\mu}G(P-\frac{q}{2})\right]\nonumber\\
&&=(-)\int\frac{d^4k}{(2\pi)^4}\int \frac{d^4l}{(2\pi)^4}~Tr_{\gamma C}\left[\gamma_{\mu}G(k)\gamma_{\mu}G(l)\right]\nonumber\\
&&=(-)\int\frac{d^4k}{(2\pi)^4}\int \frac{d^4l}{(2\pi)^4}~\frac{Tr_{\gamma C}\left\{\gamma_{\mu}[-i\gamma\cdot kA(k^2)+B(k^2)]\gamma_{\mu}[-i\gamma\cdot lA(l^2)+B(l^2)]\right\}}{[A^2(k^2)k^2+B^2(k^2)][A^2(l^2)l^2+B^2(l^2)]}\nonumber\\
&&=-48\left[\int\frac{d^4k}{(2\pi)^4}\frac{B(k^2)}{A^2(k^2)k^2+B^2(k^2)}\right]^2-\int\frac{d^4k}{(2\pi)^4}\int \frac{d^4l}{(2\pi)^4}\frac{24k\cdot lA(k^2)A(l^2)}{[A^2(k^2)k^2+B^2(k^2)][A^2(l^2)l^2+B^2(l^2)]}\nonumber\\
&&=-\frac{1}{3}\left[\int\frac{d^4k}{(2\pi)^4}~Tr_{\gamma C}\left[G(k)\right]\right]^2=-\frac{1}{3}\left[\langle\tilde{0}|:\bar{q}q:|\tilde{0}\rangle\right]^2,
\end{eqnarray}
where we have used the fact that the integrand of the second term on the fourth line is an odd function of the momenta and
hence this term vanishes. In addition, we have
\begin{eqnarray}
&&(-)\int\frac{d^4q}{(2\pi)^4}\int \frac{d^4P}{(2\pi)^4}~Tr_{\gamma C}\left[\gamma_{\mu}G^{pert}(P+\frac{q}{2})\gamma_{\mu}G^{pert}(P-\frac{q}{2})\right]=0.
\end{eqnarray}
Putting Eqs. (25) and (26) into Eq. (21), it is easy to find that vacuum saturation assumption, i.e., factorization of the four-quark condensate holds in the case of bare vertex approximation.

Similarly, for the case of the four-quark condensate
$\langle\tilde{0}|:\bar{q}(0)q(0)\bar{q}(0)q(0):|\tilde{0}\rangle$, if one adopts the bare vertex approximation 
$\Gamma(P;q) \approx I, \Gamma^{pert}(P;q) \approx I$ in Eq. (23), one can also show that factorization of $\langle\tilde{0}|:\bar{q}(0)q(0)\bar{q}(0)q(0):|\tilde{0}\rangle$ holds.  
For the case of other types of four-quark condensate:
$\langle\tilde{0}|:\bar{q}(0)\sigma_{\mu\nu}q(0)
\bar{q}(0)\sigma_{\mu\nu}q(0):|\tilde{0}\rangle$,
$\langle\tilde{0}|:\bar{q}(0)\gamma_{\mu}\gamma_5q(0)
\bar{q}(0)\gamma_{\mu}\gamma_5q(0):|\tilde{0}\rangle$
and $\langle\tilde{0}|:\bar{q}(0)\gamma_5q(0)
\bar{q}(0)\gamma_5q(0):|\tilde{0}\rangle$,
one can also prove that factorization holds in the case of
bare vertex approximation. 

From the above results, we reach a general conclusion: 
by means of the background field method, one can derive
general formula for all five types of four-quark condensates, which contains the corresponding
type of fully dressed vertex, and factorization of the four-quark condensate holds
only when the dressed vertex is taken to be the bare one.

To summarize, by differentiating the dressed quark propagator with respect to a variable background field, the linear response of the dressed quark propagator in the presence of the background field can be obtained. From this general method, using the vector background field as an illustration, we derive a general formula (21) for the four-quark condensate $\langle\tilde{0}|:\bar{q}(0)\gamma_{\mu}q(0)\bar{q}(0)\gamma_{\mu}q(0):|\tilde{0}\rangle$, which contains the corresponding dressed vertex $\Gamma_{\mu}(q;P)$ and $\Gamma_{\mu}^{pert}(q;P)$. In the case of bare vertex approximation, it is shown that factorization of the four-quark condensate holds exactly. Finally, we want to stress that the background field approach adopted in this paper is general enough and can be applied to the calculation of the vacuum expectation value of the T-product of any four quark field operators. Especially, it can be used to study the two-point correlation functions of the mesonic current,
which play an important roles in nonperturbative QCD and hadronic physics [21].

\vspace*{0.8 cm}

\noindent{\large \bf Acknowledgments}

This work was supported in part by the National Natural Science Foundation of China (under Grant Nos 10175033, 10135030 and 10575050) and the Research Fund for the Doctoral Program of Higher Education (under Grant No 20030284009).

\vspace*{0.8 cm}

\noindent{\large \bf References}
\begin{description}
\item{[1]} M. Shifman, A. Vainshtein, V. Zakharov, Nucl. Phys. {\bf B147}, 385 (1979).
\item{[2]} L. Reinders, H. Rubinstein, S. Yazaki, Phys. Rep. {\bf 127}, 1 (1985); S. Narison, QCD Spectral Sum Rules, World Scientific, Singapore, 1989, and references therein.
\item{[3]} S. Narison, R. Tarrach, Phys. Lett. {\bf B 125}, 217 (1983).
\item{[4]} G. Launer, S. Narison, R. Tarrach, Z. Phys. C. {\bf 26}, 433 (1984).
\item{[5]} R. A. Bertlmann, C. A. Dominguez, M. Loewe, M. Perrottet, E. de Rafael, Z. Phys. C {\bf 39}, 231 (1988).
\item{[6]} C. A. Dominguez, J. Sola, Z. Phys. C. {\bf 40}, 63 (1988).
\item{[7]} V. Gimenez, J. Bordes, J. Penarrocha, Nucl. Phys. B {\bf 357}, 3 (1991).
\item{[8]} D. B. Leinweber, Ann. Phys. {\bf 254}, 328 (1997).
\item{[9]} F. Klingl. N. Kaiser, W. Weise, Nucl. Phys. A {\bf 624}, 527 (1997).
\item{[10]} S. Leupold, W. Peters, U. Mosel, Nucl. Phys. A {\bf 628}, 311 (1998).
\item{[11]} S. Leupold, Phys. Lett. {\bf B 616}, 203 (2005).
\item{[12]} H. S. Zong, J. L. Ping, H. T. Yang, X. F. L\"{u}, and F. Wang, Phys. Rev. D {\bf 67}, 074004 (2003).
\item{[13]} H. S. Zong, S. Qi, W. Chen, W. M. Sun, and E. G. Zhao, Phys. Lett. {\bf B 576}, 289 (2003).
\item{[14]} R. Pascual and R. Tarrach, Renormalisation for the Practitioner (Springer-Verlag, Berlin, 1984). pp. 67-70.
\item{[15]} H. S. Zong, F. Y. Hou, W. M. Sun, J. L. Ping, and E. G. Zhao, Phys. Rev. C {\bf 72}, 035202 (2005).
\item{[16]} C. D. Roberts and A. G. Williams, Prog. Part. Nucl. Phys. {\bf 33}, 477 (1994), and references therein.
\item{[17]} C. D. Roberts and S. M. Schmidt, Prog. Part. Nucl. Phys. {\bf 45S1}, 1 (2000), and references therein.
\item{[18]} P. C. Tandy, Prog. Part. Nucl. Phys. {\bf 39}, 117 (1997); R. T. Cahill and S. M. Gunner, Fiz. {\bf B7}, 17 (1998), and references therein.
\item{[19]} H. S. Zong, Y. M. Shi, W. M. Sun, J. L. Ping, Phys. Rev. C {\bf 73}, 035206 (2006).
\item{[20]}  Y. M. Shi, K. P. Wu, W. M. Sun, H. S. Zong, and J. L. Ping, Phys. Lett. {\bf B 639}, 248 (2006).
\item{[21]} E. V. Shuryak, Rev. Mod. Phys, {\bf 65}, 1 (1993).
\end{description}

\end{document}